\documentclass{article}
\usepackage{opex3}
\usepackage{amsmath}
\usepackage{amssymb}
\usepackage{graphicx}
\usepackage{color}
\usepackage{xcolor}
\usepackage{cite}

\begin{document}

\title{$PT$ symmetry breaking and nonlinear optical isolation in coupled microcavities}

\author{Xin Zhou$^{1}$ and Y.~D.~Chong$^{1,2,*}$}

\address{$^1$ Division of Physics and Applied Physics, School of Physical and Mathematical Sciences, Nanyang Technological University, Singapore 637371, Singapore \\
$^2$ Centre for Disruptive Photonic Technologies, Nanyang Technological University, Singapore 637371, Singapore}

\email{$^*$ yidong@ntu.edu.sg}

\begin{abstract}
We perform a theoretical study of the nonlinear dynamics of nonlinear
optical isolator devices based on coupled microcavities with gain and
loss. This reveals a correspondence between the boundary of asymptotic
stability in the nonlinear regime, where gain saturation is present,
and the $PT$-breaking transition in the underlying linear system.  For
zero detuning and weak input intensity, the onset of optical isolation
can be rigorously derived, and corresponds precisely to the transition
into the $PT$-broken phase of the linear system.  When the couplings
to the external ports are unequal, the isolation ratio exhibits an
abrupt jump at the transition point, whose magnitude is given by the
ratio of the couplings.  This phenomenon could be exploited to realize
an actively controlled nonlinear optical isolator, in which strong
optical isolation can be turned on and off by tiny variations in the
inter-resonator separation.
\end{abstract}


\section{Introduction}
\label{introduction}

For many years, the implementation of compact optical isolators has
been a major research goal in the field of integrated optics
\cite{soljacic_review,Dirk2013}.  Optical isolation requires the
breaking of Lorentz reciprocity; this is traditionally achieved using
magneto-optic materials, but such materials are challenging to
incorporate into integrated optics devices \cite{JOSAB1, JOSAB2}.  The
most commonly-pursued alternative method for breaking reciprocity is
to exploit optical nonlinearity \cite{soljacic_review, Gallo2001,
  Miroschnichenko2010, Krause2008, Poulton2010, nature1,
  nature2,Fan2015}.  Two recent demonstrations of nonlinearity-based
on-chip optical isolators, by Peng \textit{et al.}~\cite{nature1} and
Chang \textit{et al.}~\cite{nature2}, have drawn particular attention.
These experiments featured a pair of coupled whispering-gallery
microcavities, one containing loss and the other saturable (nonlinear)
gain.  Light transmission across the structure was found to be
strongly nonreciprocal, depending on whether it first passed through
the gain or loss resonator.  Aided by the high $Q$ factors of the
resonators, isolation was observed for record-low powers of $\sim
1\mu$W \cite{nature1}.

The use of dual resonators containing gain and loss in \cite{nature1,
  nature2} was inspired by ``$PT$ symmetric optics'', which concerns
optical structures that are invariant under simultaneous parity-flip
($P$) and time-reversal ($T$) operations
\cite{ElGanainy,Kostas1,Kostas2,Musslimani1,
  Musslimani2, Guo,Ruter,Peschel,Feng2013,Longhi2010,Chong2011,Wimmer2015}.  The
concept originated from the observation that $PT$ symmetric
Hamiltonians, despite being non-Hermitian, can exhibit real eigenvalue
spectra \cite{Bender,Bender02}, as well as ``$PT$-breaking
transitions'' between real and complex eigenvalue regimes.  The
$PT$-breaking transition point is an ``exceptional point'', where two
eigenstates coalesce and the effective Hamiltonian becomes defective
\cite{exceptional_points,Ramezani2012}.  Near the transition, the
dynamical behavior of the optical fields can exhibit highly
interesting features \cite{Lu2015,Kepesidis2015,Hassan2015,Peng2014};
for instance, the presence of gain saturation has been found to
stabilize $PT$-symmetric steady states past the usual $PT$ transition
point \cite{Kepesidis2015,Hassan2015}.

Despite these intriguing conceptual links, it was not clear from
\cite{nature1, nature2} how $PT$ symmetry relates to the working of
the nonlinear optical isolators in question.  Strictly speaking, $PT$
symmetry holds in the dual-resonator structures only in the linear
limit; in the nonlinear regime, the gain saturates and no longer
matches the loss, so the structures are not $PT$ symmetric and do not
possess distinct ``$PT$-symmetric'' or ``$PT$-broken'' phases.  Peng
\textit{et al.}, in \cite{nature1}, indicated that optical isolation
occurs (in the nonlinear regime) if the system is tuned so that it
would be $PT$-broken \textit{in the linear regime}; however, the
actual correspondence was not shown theoretically nor experimentally.
The dynamical behavior of the system, including the uniqueness and
stability of the steady-state solution(s), was also unexplored.

In this paper, we present a theoretical analysis of the dual-resonator
structure, aiming to clarify the relationships between the $PT$ phase,
the performance of the nonlinear optical isolator, and the uniqueness
and stability of the steady-state optical modes.  Using coupled-mode
theory \cite{CMT1,CMT2,CMT3,CMT4}, we study the conditions for
steady-state solutions to exist, and the asymptotic stability of those
solution(s).  We find that stability in the nonlinear system has a
close correspondence with the $PT$ transition boundary of the
underlying linear system.

In the ``weak-input limit'', where the input intensity is low relative
to the gain saturation threshold within the amplifying resonator, we
show that the nonlinear solutions at non-zero frequency detunings are
asymptotically stable in the $PT$-symmetric phase.  In the $PT$-broken
phase, the solutions become unstable at sufficiently large frequency
detunings, and the nonlinear system exhibits limit-cycle oscillations,
which might be useful for frequency generation applications (such as
frequency combs).

For small frequency detunings, multiple steady-state solutions can
exist in the $PT$-broken phase, but only the highest-intensity
solution is asymptotically stable.  Specifically at zero detuning,
there is always one stable steady-state solution, and the
\textit{nonlinear} system exhibits a sharp transition between
isolating behavior (corresponding to the $PT$-broken phase) and
reciprocal behavior (corresponding to the $PT$-symmetric phase).
Although this transition coincides exactly with the $PT$ transition
point, it is an inherently nonlinear effect, arising from a jump
between different solution branches of the transmission intensity
equations.  However, the performance of the isolator can be
significantly limited by the contributions to the nonlinearity caused
by a reflected wave \cite{Fan2015}.

We also show that the performance of the nonlinear optical isolator is
also modified in a useful way when the two resonator-to-waveguide
coupling rates are unequal.  In this case, a small shift across the
transition point causes the isolation ratio (the ratio between forward
and backward transmission intensities) to undergo an abrupt jump,
which approaches a discontinuity in the weak-input limit.  The
magnitude of this jump is given by the ratio of the coupling rates.
This phenomenon can be used to realize a nonlinear optical isolator
that exhibits very large changes in the isolation ratio, actively
controlled by tiny shifts in (e.g.) the inter-resonator separation.

\section{Coupled-mode equations}

\begin{figure}
  \centering\includegraphics[width=0.9\textwidth]{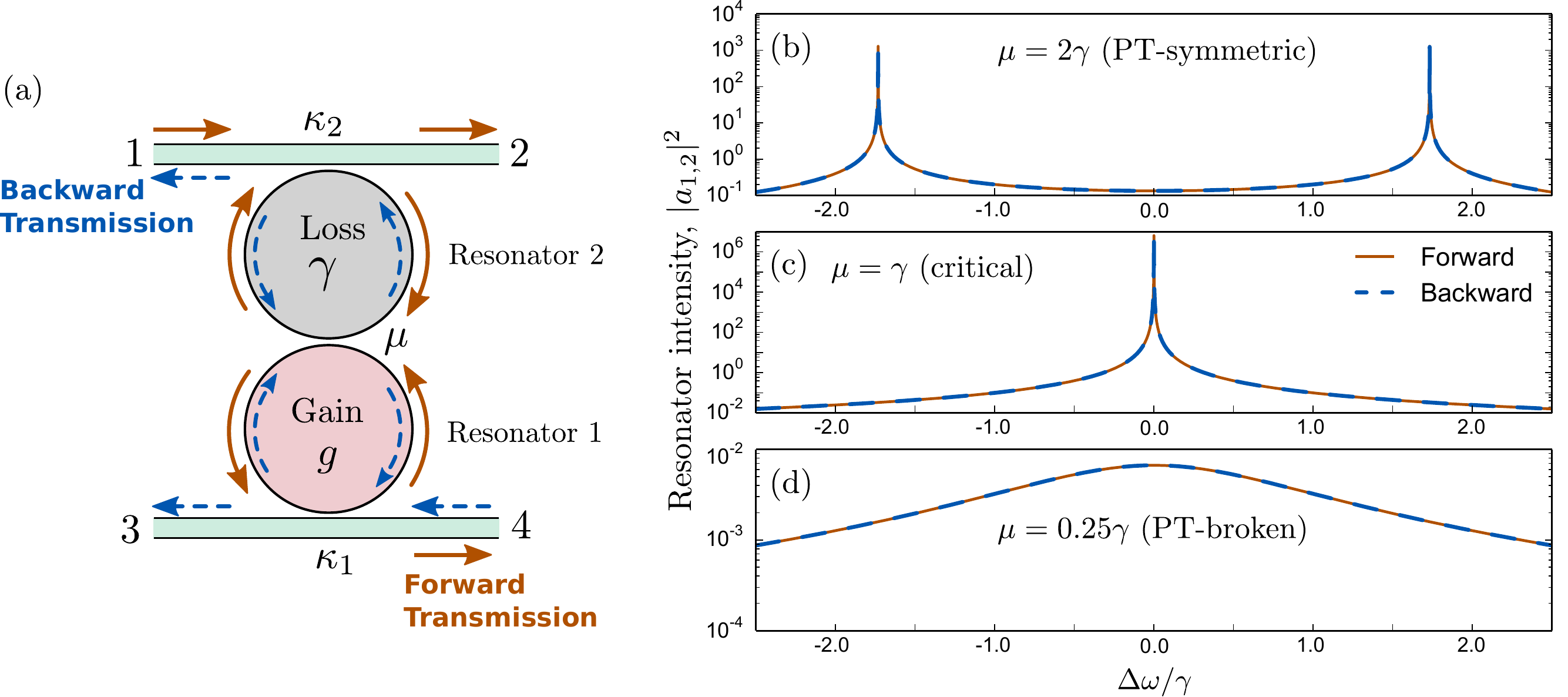}
  \caption{(a) Schematic of a resonator with saturable gain coupled to
    a lossy resonator, with both resonators coupled to optical fiber
    ports.  Solid arrows indicate forward transmission (port
    $1\rightarrow 4$), and dashed arrows indicate backward
    transmission (port $4\rightarrow 1$).  (b)--(d) Transmission
    characteristics in the linear (non-gain-saturated) regime, when
    the gain and loss are $PT$ symmetric ($g = \gamma = 0.4$).  Here,
    we plot the intensity in the active resonator ($|a_1|^2$) under
    forward transmission (solid lines), and in the passive resonator
    ($|a_2|^2$) under backward transmission (dashes), versus the
    frequency detuning.  These resonator intensities are proportional
    to the forward and backward transmission intensities via
    Eqs.~(\ref{forward3}) and (\ref{back3}).  In the $PT$-symmetric
    phase $\mu > \gamma$, there are two transmission peaks; in the
    $PT$-broken phase $\mu < \gamma$, these merge into a single peak.}
  \label{fig:schematic}
\end{figure}

The dual-resonator structure is shown schematically in
Fig.~\ref{fig:schematic}(a).  The setup is identical to the
experiments reported in \cite{nature1,nature2}, consisting of two
evanescently coupled microcavities with resonant frequencies
$\omega_1$ and $\omega_2$.  One resonator contains saturable gain, and
the other is lossy.  The resonators are coupled to separate optical
fiber waveguides, which act as input/output ports (labeled 1--4), with
couplings $\kappa_1$ and $\kappa_2$.  The direct inter-resonator
coupling rate is $\mu$.  In the ``forward transmission''
configuration, light is injected from port 1 at a fixed operating
frequency $\omega$, exiting at ports 2 and 4.  Alternatively, in the
``backward transmission'' configuration, light is injected at port 4
and exit at ports 1 and 3.  We are interested in the level of
isolation between ports 1 and 4, which serve as the operational input
and output ports for the device.

The dual-resonator system can be described by coupled-mode equations
\cite{nature1,nature2}, formulated using the standard framework of
coupled-mode theory \cite{CMT1,CMT2,CMT3,CMT4}.  In this section and
the next, we briefly summarize these equations, which have previously
been presented in~\cite{nature1,nature2}.  For forward
transmission, the coupled-mode equations are
\begin{align}
  \frac{da_1}{dt} &= (i\Delta\omega_1+g) a_1 - i\mu a_2 \label{forward1} \\ 
  \frac{da_2}{dt} &= (i\Delta\omega_2-\gamma) a_2 - i\mu a_1 + \sqrt{\kappa_2}s_{\mathrm{in}} 
  \label{forward2} \\
  I_F &= \kappa_1 |a_1|^2. \label{forward3}
\end{align}
Here, $a_{1}$ and $a_{2}$ denote the complex amplitudes for the
slowly-varying field amplitudes in the gain resonator and loss
resonators, respectively; $\Delta \omega_{1,2} \equiv \omega -
\omega_{1,2}$ denote the operating frequency's detuning from each
resonator's natural frequency; $g > 0$ and $\gamma > 0$ are the net
gain rate in resonator 1 and the net loss rate in resonator 2;
$s_{\mathrm{in}}$ is the amplitude of the incoming light in port 1;
and $I_F$ is the power transmitted forward into port 4.  For the
moment, we assume that there is no reflected wave re-entering the
system from port 4; the effects of such a reflected wave will be
discussed in Section \ref{Effect of a Back-Reflected Wave}.

For backward transmission, a different set of coupled-mode equations
holds:
\begin{align}
  \frac{da_1}{dt} &= (i\Delta\omega_1 + g) a_1 - i \mu a_2
  + \sqrt{\kappa_1}\, s_{\mathrm{in}} \label{back1} \\
  \frac{da_2}{dt} &= (i\Delta\omega_2-\gamma)a_2 - i\mu a_1 \label{back2} \\
  I_B &= \kappa_2 \, |a_2|^2, \label{back3}
\end{align}
where $I_B$ is the power transmitted into port 1.

The gain/loss rates $g$ and $\gamma$ consist of several radiative and
non-radiative terms \cite{nature1}:
\begin{align}
  g &= \frac{1}{2}\left(g' - \gamma_1 - \kappa_1\right) \\
  \gamma &= \frac{1}{2}\left(\gamma_2 + \kappa_2\right),
\end{align}
where $g'$ is the intrinsic amplification rate in resonator 1, and
$\gamma_{1,2}$ are the intrinsic loss rates in the resonators.  Until
stated otherwise, we will impose the following simplifying
restrictions:
\begin{align}
  \kappa_1 &= \kappa_2 = \gamma_1 = \gamma_2, \label{equal_loss} \\
  \Delta \omega_1 &= \Delta \omega_2 \equiv \Delta \omega, \label{tuning} \\
  g' &= \frac{g_0}{1+\left|a_1/a_s\right|^{2}}. \label{gain}
\end{align}
Equation~(\ref{equal_loss}) corresponds to a ``critical coupling''
criterion with respect to the individual cavity-waveguide couplings.
The intrinsic loss and outcoupling rates are all tuned to the same
value; note also that $g = g'/2 - \gamma$.  Equation~(\ref{tuning}) states
that the resonators have the same natural frequency.
Equations~(\ref{equal_loss})--(\ref{tuning}) serve as simplifying
assumptions, to avoid dealing with a proliferation of free parameters;
later, we will discuss the implications of relaxing these assumptions.
Another important constraint, $PT$ symmetry, will be imposed in the
next section.  Equation~(\ref{gain}) describes saturable gain, where $g_0$
is the unsaturated amplification rate, and $a_s \in \mathbb{R}^+$ is a
gain saturation threshold.

The experimentally realized systems reported in \cite{nature1,nature2}
operated in the 1550 nm wavelength band, with rate parameters $\mu$,
$g_0$, $\gamma$ and $\kappa_{1,2}$ on the order of 10 MHz in
\cite{nature1}, and 100 MHz in \cite{nature2}.  The coupling rates
$\mu$ and $\kappa_{1,2}$ can be tuned via the inter-resonator and
resonator-waveguide separations.  The input power
$|s_{\mathrm{in}}|^2$ ranged from zero to around 10--100
$\mu$W~\cite{nature1,nature2}.

The suitability of the system as an optical isolator is characterized
using the ``isolation ratio'', which is the ratio of forward to
backward transmittance at fixed input power:
\begin{equation}
  R \equiv \frac{T_F}{T_B} = \frac{I_F(I_{\mathrm{in}})}{I_B(I_{\mathrm{in}})},
  \label{isolation_ratio}
\end{equation}
where $I_F$ is obtained by solving
Eqs.~(\ref{forward1})--(\ref{forward3}) with $\dot{a_1} = \dot{a_2} =
0$ (steady state), and $I_B$ is obtained from
Eqs.~(\ref{back1})--(\ref{back3}).  When the system is reciprocal,
$R=1$.  The isolation ratio was also used in \cite{nature1,nature2} as
the figure of merit for optical isolation.  However, it is worth
noting that the forward and backward transmissions are being compared
under the assumption that, in either case, no reflected wave is
present.  We will discuss this limitation in greater detail in Section
\ref{Effect of a Back-Reflected Wave}.

\section{Linear operation}
\label{sec:linear}

We now impose the important constraint $g = \gamma$.  This means that
in the linear regime, $a_s \rightarrow \infty$, the gain and loss
resonators become $PT$ symmetric.  To understand the implications,
consider the ``closed'' system without resonator-fiber couplings.  Its
detuning eigenfrequencies are
\begin{eqnarray}
  \Delta \omega = i\,\frac{g-\gamma}{2}
  \pm \sqrt{\mu^2 - \gamma g - \left(\frac{g-\gamma}{2}\right)^2}.
\end{eqnarray}
When $g = \gamma$, these reduce to $\Delta \omega = \pm \sqrt{\mu^2 -
  \gamma^2}$.  As $\mu$ and $\gamma$ are varied while keeping $g =
\gamma$, the system has a $PT$ symmetry-breaking transition at $\mu =
\gamma$.  For $\mu > \gamma$, the detunings are real ($PT$-symmetric
phase), and for $\mu < \gamma$ they are purely imaginary ($PT$-broken
phase).

With the resonator-fiber couplings introduced, the eigenmodes become
transmission resonances.  In the $PT$-symmetric phase $\mu > \gamma$,
the resonator modes and transmission amplitudes exhibit two intensity
peaks, at $\Delta \omega = \pm \sqrt{\mu^2 - \gamma^2}$, corresponding
to the (real) detunings of the closed system, as shown in
Fig.~\ref{fig:schematic}(b).  In the $PT$-broken phase $\mu < \gamma$,
there is a single peak at zero detuning, as shown in
Fig.~\ref{fig:schematic}(d).  As noted by Peng \textit{et
  al.}~\cite{nature1}, the $PT$-symmetric and $PT$-broken phases will
give very different behaviors once gain saturation is introduced.

In the linear regime, Eqs.~(\ref{forward1})--(\ref{forward3}) and
Eqs.~(\ref{back1})--(\ref{back3}) obey optical reciprocity by explicit
construction \cite{CMT2}.  For fixed $s_{\mathrm{in}}$, the forward
and backward transmission amplitudes are exactly equal, $I_F = I_B$;
the isolation ratio is $R = 1$, as shown in
Fig.~\ref{fig:schematic}(b)--(d).

\section{Nonlinear operation: multiple solutions and stability}
\label{sec:multisols}

We turn now to the nonlinear, gain-saturated regime, setting $g_0 =
4\gamma$, so that $g \rightarrow \gamma$ as $a_s \rightarrow \infty$.
This means the system would be $PT$ symmetric in the absence of gain
saturation.  If we use $a_s$ as the natural intensity scale for the
coupled-mode equations, the nonlinear system has four remaining
independent parameters: $\Delta \omega$, $\mu$, $\gamma$, and
$|s_{\mathrm{in}}|^2$.


For finite $a_s$, optical reciprocity is broken.  However, the system
is no longer $PT$ symmetric, since $g \ne \gamma$, and thus we can no
longer rigorously define ``$PT$ symmetric'' or ``$PT$ broken'' phases.
Still, we can relate the nonlinear system's behavior to the $PT$
symmetric phases \textit{as defined in the linear limit}.

\begin{figure}
  \centering\includegraphics[width=0.98\textwidth]{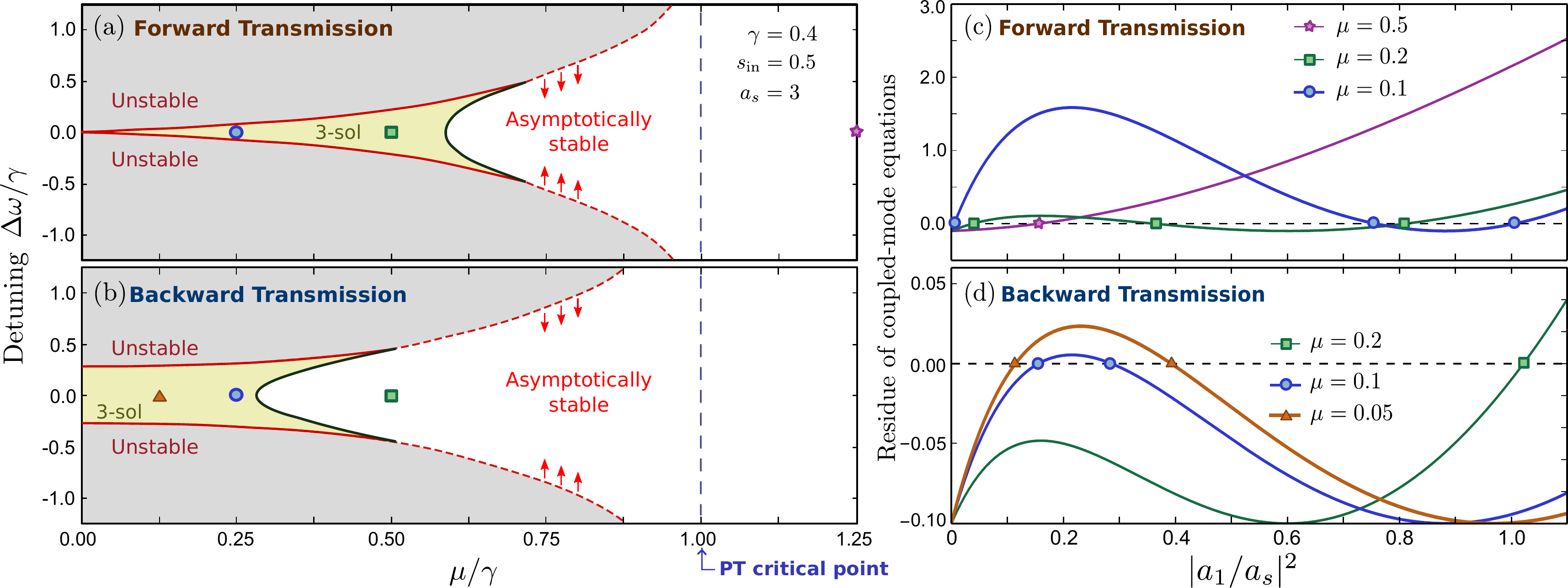}
  \caption{(a)--(b) Domains in which the nonlinear coupled-mode
    equations have multiple steady-state solutions, for forward (a)
    and backward (b) transmission.  Here, we show the parameter space
    defined by the frequency detuning $\Delta \omega$ and
    inter-resonator coupling $\mu$, with fixed $\gamma=0.4$,
    $s_{\mathrm{in}} = 0.5$, and $a_s = 3$; symbols indicate the
    points in the parameter space corresponding to the curves in (c)
    and (d).  Within the small-$\Delta\omega$ region bounded by the
    red curves, the highest-intensity (or only) solution is
    asymptotically stable.  (c)--(d) Plots showing the emergence of
    multiple solutions at several values of $\mu$, fixing $\Delta
    \omega = 0$.  The horizontal axis is the normalized intensity in
    the gain resonator, $|a_1/a_s|^2$; the vertical axis is the
    left-hand side of the cubic Eq.~(\ref{cubic}), and its counterpart
    for backward transmission; the steady-state coupled-mode equations
    are satisfied when the curves cross zero.  }
  \label{fig:multisolutions}
\end{figure}

In the linear regime, the solutions to the coupled-mode equations were
unique.  With nonlinearity, the coupled-mode equations can have
multiple steady-state solutions.  For forward transmission,
steady-state solutions are determined by combining
Eqs.~(\ref{forward1})--(\ref{forward2}) into:
\begin{equation}
  |\alpha|^2 x^3 +
  \Big(2|\alpha - 1|^2 - 2
    - \beta \Big) x^2
  + \Big(|\alpha-2|^2 - 2 \beta\Big) x - \beta = 0,
  \label{cubic}
\end{equation}
where $\alpha$, $\beta$, and $x$ are the following dimensionless
variables:
\begin{equation}
  \alpha = \frac{(-i\Delta\omega + \gamma)^2 + \mu^2}
         {\gamma(-i\Delta\omega + \gamma)},\quad
  \beta = \frac{\mu^2}{\Delta\omega^2 + \gamma^2}\, \frac{1}{\gamma}
  \, \left|\frac{s_{\mathrm{in}}}{a_s}\right|^2,\quad
  x = \left|\frac{a_1}{a_s}\right|^2.
  \label{alphabeta}
\end{equation}
Since $x \in \mathbb{R}^+$, there is either one, two, or three
physical steady-state solutions.  There must be at least one solution,
since the polynomial has a positive third-order coefficient and
negative zeroth-order coefficient.  The backward transmission case is
handled similarly, using Eqs.~(\ref{back1})--(\ref{back2}); it gives
the same cubic equation as Eq.~(\ref{cubic}), but with the replacement
\begin{equation}
  \beta = \frac{1}{\gamma}\, \left|\frac{s_{\mathrm{in}}}{a_s}\right|^2.
  \label{beta_back}
\end{equation}

Solving the polynomial reveals a domain in parameter space where there
are three physical steady-state solutions, outside of which the
solution is unique.  This is shown in
Fig.~\ref{fig:multisolutions}(a)--(b).  The three-solution domain lies
within the ``$PT$-broken'' phase of the linear system, $\mu < \gamma$.

The boundaries of the three-solution domain depend on $\gamma$ and
$s_{\mathrm{in}}$, as the choice of forward or backward transmission.
It consists of two sets of curves; the black curves in
Fig.~\ref{fig:multisolutions}(a)--(b) involve a degeneracy of two
low-intensity roots of the cubic polynomial
[Fig.~\ref{fig:multisolutions}(c)--(d)].  Crossing this boundary
causes no discontinuity in the intensity of the stable steady-state
solution.  The red curves in Fig.~\ref{fig:multisolutions}(a)--(b)
involve the degeneracy of two high-intensity roots of the cubic
polynomial (\ref{cubic}); crossing this boundary destabilizes the
steady-state solution.

Through numerical stability analysis, detailed in Appendix A, we find
that the highest-intensity solution in the three-solution domain is
asymptotically stable (i.e., the Lyapunov exponents are all negative).
The two lower-intensity solutions are \textit{unstable}: small
perturbations from these steady states eventually evolve into the
highest-intensity state.  In the one-solution domain, the solution is
asymptotically stable for small detuning $\Delta \omega$, and unstable
for large $\Delta \omega$.

Interestingly, the region of asymptotic stability in the nonlinear
system is closely connected to the $PT$ symmetry phases of the linear
system.  For $\mu < \gamma$, which corresponds to the $PT$-broken
phase, the frequency range of asymptotic stability is bounded by the
solid and dashed red curves shown in
Fig.~\ref{fig:multisolutions}(a)--(b).  These bounds diverge at $\mu =
\gamma$, which corresponds to the transition from the $PT$-broken to
the $PT$-symmetric phase in the linear system.  For $\mu > \gamma$,
the steady state solution becomes asymptotically stable for all
$\Delta \omega$.

In the one-solution domain, the onset of asymptotic instability (at
sufficiently large $\Delta \omega$) is associated with the appearance
of sustained time-domain beating in both the resonator intensities and
the transmittance.  This is a Hopf bifurcation \cite{hopf_text} from a
stable state to limit cycle behavior (see Fig.~\ref{fig:numerical} in
Appendix A).


\section{Isolation ratios at zero detuning}
\label{sec:Isolation Ratios at Zero Detuning}

Let us now focus on zero detuning, $\Delta \omega = 0$.  In this case,
there is always an asymptotically stable steady-state solution, and we
shall be able to derive an important connection to the $PT$ transition
of the linear system.  The variables $\alpha$ and $\beta$, defined in
Eq.~(\ref{alphabeta}), simplify to
\begin{align}
  \alpha &= 1 + \left(\frac{\mu}{\gamma}\right)^2 \\
  \beta &= \frac{\left|s_{\mathrm{in}}/a_s\right|^2}{\gamma} \times
  \left\{\begin{array}{ll}\displaystyle\left(\frac{\mu}{\gamma}\right)^2,
  & (\text{Forward}) \\
  1, &(\text{Backward}).
  \end{array}\right.
  \label{beta_omega0}
\end{align}
Hence, the cubic polynomial in Eq.~(\ref{cubic}) is entirely
determined by two quantities: (i) $\mu/\gamma$ and (ii)
$|s_{\mathrm{in}}/a_s|^2/\gamma$.  The first quantity is also the
tuning parameter for the $PT$ transition.  The second quantity
determines the strength of the input relative to the gain saturation
threshold.  We will be particularly interested in the ``weak-input''
limit, defined as
\begin{equation}
  s_{\mathrm{in}} \ll \sqrt{\gamma}\, a_s.
  \label{weakinput}
\end{equation}
When $\beta \ll 1$, the steady state behavior will be principally
determined by the $PT$-tuning parameter $\mu/\gamma$.

\begin{figure}
  \centering\includegraphics[width=0.9\textwidth]{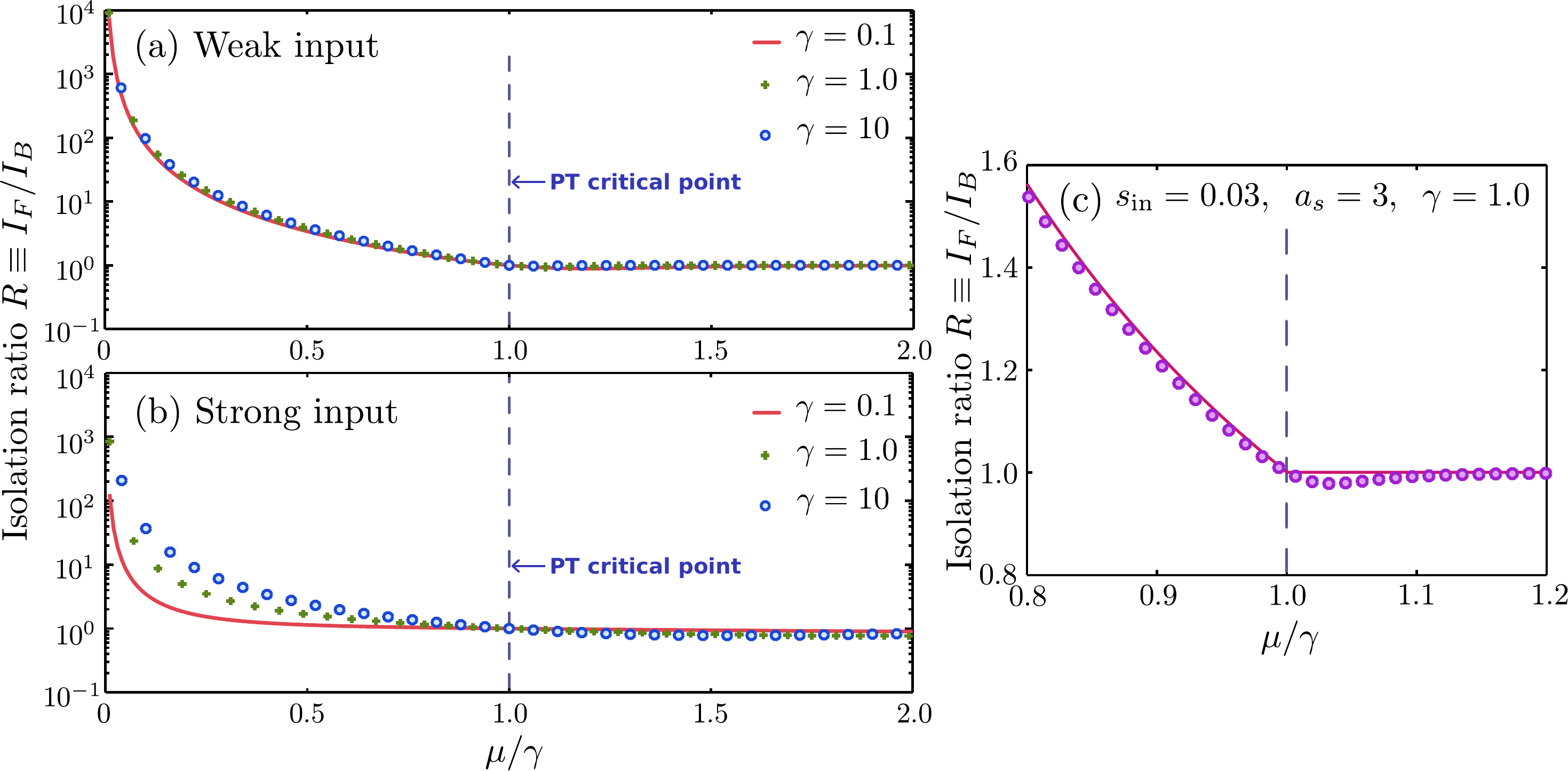}
  \caption{Isolation ratio versus $\mu/\gamma$ at zero frequency
    detuning ($\Delta \omega = 0$), for (a) weak inputs
    $s_{\mathrm{in}} = 0.15$ and $a_s = 3$, and (b) strong inputs
    regime $s_{\mathrm{in}} = 9$ and $a_s = 3$, using several choices
    of $\gamma$.  In the weak-input regime, the isolation ratio is
    mainly determined by the $PT$-breaking parameter $\mu/\gamma$.
    The system becomes reciprocal for $\mu/\gamma > 1$, corresponding
    to the $PT$-symmetric phase of the linear system.  (c) Close-up of
    the isolation ratio behavior in the weak-input regime, showing the
    kink in the dependence on $\mu/\gamma$ at the $PT$ transition
    point $\mu/\gamma = 1$.  Circles show exact numerical solutions of
    the coupled-mode equations, and the solid curve shows the analytic
    approximations of Eqs.~(\ref{R0})--(\ref{R1}). }
  \label{fig:rescale_high_low}
\end{figure}

Figure~\ref{fig:rescale_high_low} plots the isolation ratio $R \equiv
I_F/I_B$ versus $\mu / \gamma$, for several different values of
$\gamma$ and $s_{\mathrm{in}}$.  In the weak-input regime, the
isolation ratio curves are almost identical for different $\gamma$,
which verifies that the system is controlled by the combination
$\mu/\gamma$.  For $\mu/\gamma < 1$, corresponding to the $PT$-broken
phase of the linear system, we find that $R > 1$, and hence the system
functions as a good optical isolator.  For $\mu/\gamma > 1$, we find
that $R \approx 1$.  This agrees with the qualitative behaviors
reported in \cite{nature1}.

Let us examine the vicinity of the transition point in greater detail.
Figure~\ref{fig:rescale_high_low}(c) shows that in the weak input
regime, the isolation ratio curve exhibits a kink at $\mu/\gamma = 1$.
To understand this, we return to the definition of the isolation
ratio:
\begin{equation}
  R = \frac{I_F}{I_B} = \left(\mu/\gamma\right)^{-2} \, \frac{x_F}{x_B},
  \label{Romega0}
\end{equation}
where $x_F$ and $x_B$ are the solutions to Eq.~(\ref{cubic}) for the
forward and backward transmission cases.  For $\beta \rightarrow 0$,
Eq.~(\ref{cubic}) reduces to
\begin{equation}
  x\left(x - \frac{1-(\mu/\gamma)^2}{1+(\mu/\gamma)^2}\right)^2 \approx 0.
  \label{simplecubic}
\end{equation}
For $\mu/\gamma < 1$, the double-root in Eq.~(\ref{simplecubic}) is
positive.  Hence, in this approximation, the three-solution domain
discussed in Section \ref{sec:multisols} extends over the entire range
$\mu/\gamma < 1$ along the zero-detuning line.  The asymptotically
stable solution corresponds to the double-root, which is equal for
forward and backward transmission, to lowest order in $\beta$.  Hence,
we can use Eq.~(\ref{Romega0}) to show that
\begin{equation}
  R \approx \left(\mu/\gamma\right)^{-2} \;\;\;\text{for}\;\; \mu/\gamma < 1,
  \;\;s_{\mathrm{in}} \ll \sqrt{\gamma}\, a_s.
  \label{R0}
\end{equation}

For $\mu/\gamma > 1$, the double-root is negative, so the only valid
root in the $\beta \rightarrow 0$ limit is $x = 0$.  For non-zero
$\beta$, this root becomes $\mathcal{O}(\beta)$, so
Eq.~(\ref{beta_omega0}) implies that $x_F/x_B \approx (\mu/\gamma)^2$.
This yields the isolation ratio
\begin{equation}
  R \approx 1 \quad\quad\quad\text{for}\;\; \mu/\gamma > 1,
  \;\;s_{\mathrm{in}} \ll \sqrt{\gamma}\, a_s.
  \label{R1}
\end{equation}
The limiting expressions (\ref{R0})--(\ref{R1}) are plotted in
Fig.~\ref{fig:rescale_high_low}(c), and agree well with the numerical
solutions.  This helps explain why the $PT$ phase of the linear system
affects the isolation functionality of the nonlinear system.  Both
phenomena are determined by the parameter $\mu/\gamma$, with a
critical point at $\mu/\gamma = 1$.  The kink in the isolation ratio
arises from switching solution branches at the critical point.

\section{Imbalanced input/output couplings}

Thus far, we have assumed that the waveguide-resonator couplings,
$\kappa_1$ and $\kappa_2$, are equal.  If the couplings are unequal,
the isolation behavior of the system can be quite different.  To study
this, we replace Eq.~(\ref{equal_loss}) with
\begin{equation}
  \gamma_1 + \kappa_1 = \gamma_2 + \kappa_2 = 2\gamma.
  \label{unequal_losses}
\end{equation}
For $g_0 = 4\gamma$, the gain in resonator 1 is
\begin{equation}
  g = \frac{2\gamma}{1+|a_1/a_s|^2} - \gamma,
\end{equation}
which ensures that the decoupled system remains $PT$ symmetric with
critical point $\mu = \gamma$, as before.  With this generalization,
the steady-state equations (\ref{cubic})--(\ref{beta_back}) are
altered only by the replacements
\begin{align}
  \begin{aligned}
    \beta &\rightarrow \frac{\kappa_2}{\gamma} \beta \;\;\;(\text{Forward}) \\
    \beta &\rightarrow \frac{\kappa_1}{\gamma} \beta \;\;\;(\text{Backward}).
  \end{aligned}
  \label{beta_corrections}
\end{align}
By varying the couplings and losses so that Eq.~(\ref{unequal_losses})
is satisfied, we can access different values of $\kappa_1/\kappa_2$,
subject to the constraint $0 < \kappa_{1}, \kappa_{2} < 2\gamma$.

\begin{figure}
  \centering\includegraphics[width=0.6\textwidth]{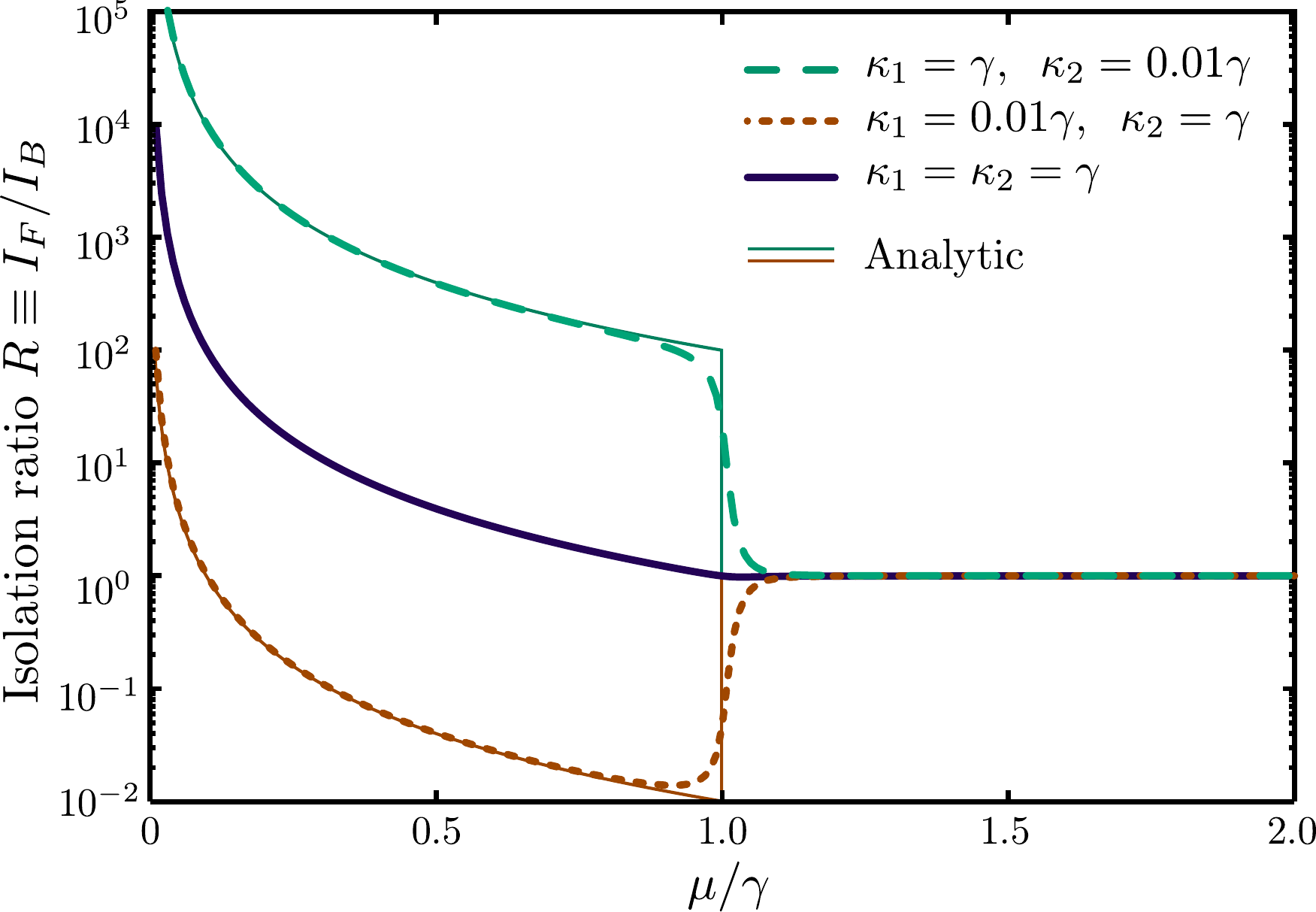}
  \caption{Isolation ratio versus $\mu/\gamma$ for different
    microcavity-waveguide coupling rates.  The system parameters are
    $\Delta \omega = 0$, $s_{\mathrm{in}} = 0.03$, $a_s = 3$, $\gamma
    = 1$, and $g_0 = 4\gamma$.  Thin solid lines show the analytic
    approximation in the weak-input limit ($s_{\mathrm{in}} \ll
    \sqrt{\gamma}\, a_s$), given by Eq.~(\ref{Runequal}).  }
  \label{fig:unequal_kappa}
\end{figure}

The discussion of Section \ref{sec:Isolation Ratios at Zero Detuning}
generalizes to this case in a straightforward way.  Using the previous
zero-detuning and weak-input assumptions, we find that $x_F/x_B
\approx 1$ for $\mu/\gamma < 1$, as before; but for $\mu/\gamma > 1$,
Eq.~(\ref{beta_corrections}) gives $x_F/x_B \approx \beta_F/\beta_B =
(\kappa_2/\kappa_1)(\mu/\gamma)^2$.  The isolation ratio now becomes
\begin{align}
  \begin{aligned}
    R &= (\kappa_1/\kappa_2) (\mu/\gamma)^{-2} x_F/x_B \\
    &\approx \left\{\begin{array}{ll}
    (\kappa_1/\kappa_2) \, \left(\mu/\gamma\right)^{-2} &\text{for}\;\; \mu/\gamma < 1 \\
    1 &\text{for}\;\; \mu/\gamma > 1.
  \end{array} \right.
  \label{Runequal}
  \end{aligned}
\end{align}
For $\kappa_1\ne\kappa_2$, this predicts a discontinuity in the
isolation ratio at $\mu/\gamma = 1$.

Figure~\ref{fig:unequal_kappa} plots dependence of the isolation ratios
on $\mu/\gamma$, at zero detuning, for the cases of (i) $\kappa_1 \ll
\kappa_2$, (ii) $\kappa_1 \gg \kappa_2$, and (iii) $\kappa_1 =
\kappa_2$.  In all three cases, the isolation ratio approaches unity
for $\mu/\gamma > 1$.  However, for $\mu/\gamma < 1$, the
unequal-coupling curves exhibit an abrupt change corresponding to a
factor of $\kappa_1/\kappa_2$ (which is two orders of magnitude for
these examples).  Interestingly, for $\kappa_1 \ll \kappa_2$, the
isolation ratio in fact \textit{decreases} below unity, before
increasing again as $\mu/\gamma \rightarrow 0$.  The numerical results
match Eq.~(\ref{Runequal}) very well.

This phenomenon may be exploited in device applications for realizing
an actively switchable optical isolator.  Using a small variation in
the $\mu/\gamma$ parameter (e.g., by varying the inter-cavity
separation, which affects $\mu$), we can switch between strong
optically isolating and reciprocal regimes.

\section{Effect of a simultaneous reflected wave}
\label{Effect of a Back-Reflected Wave}

We have analyzed the nonlinear system and its isolation ratio under
the assumption that light propagates in one direction at a time (i.e.,
forward or backward).  This is a good assumption if the isolator is
part of a optical circuit operating with optical pulses, such that any
reflected pulse re-entering the isolator (due to scattering from other
parts of the circuit) does so at a later time, after the initial pulse
has already died away.  When forward and backward waves are
simultaneously present, however, both contribute to the nonlinearity,
causing the isolator to fail.  This is a general limitation of optical
isolators based on nonlinearity \cite{Fan2015}.

In order to model simultaneous forward and backward waves, we modify
the coupled-mode equations to include resonator modes with the
opposite circulation.  Similar to the previous ``forward''
configuration, we suppose the main incident wave enters at port 1 with
amplitude $s_f$.  In addition, there is a back-propagating wave,
incident at port 4 with amplitude $s_b$, which eventually exits at
port 1. The modified equations are (taking $\Delta \omega = 0$ for
simplicity):
\begin{align}
  \frac{da_1}{dt} &= g a_1 - i\mu a_2 \\
  \frac{da_2}{dt} &= -\gamma a_2 - i\mu a_1 +
\sqrt{\kappa_2}s_f \\
  \frac{da_1'}{dt} &= g a_1' - i\mu a_2' + \sqrt{\kappa_1} s_b \label{reflectioneq} \\
  \frac{da_2'}{dt} &= -\gamma a_2' - i\mu a_1'.
\end{align}
The opposite-circulation mode amplitudes are denoted by $a_1'$ and
$a_2'$.  The back-propagating wave enters into the final term on the
right-hand side of Eq.~(\ref{reflectioneq}), coupling to the $a_1'$
mode.  Both modes in the gain resonator now contribute to the gain
saturation, so
\begin{equation}
  g = \frac{2\gamma}{1+\left|a_1/a_s\right|^{2} + \left|a_1'/a_s\right|^{2}} - \gamma.
\end{equation}
Instead of the isolation ratio, we now consider the transmittance of
the backward wave:
\begin{equation}
  T_b = \frac{\kappa_2 |a_2'|^2}{|s_b|^2}.
\end{equation}

\begin{figure}
  \centering\includegraphics[width=0.46\textwidth]{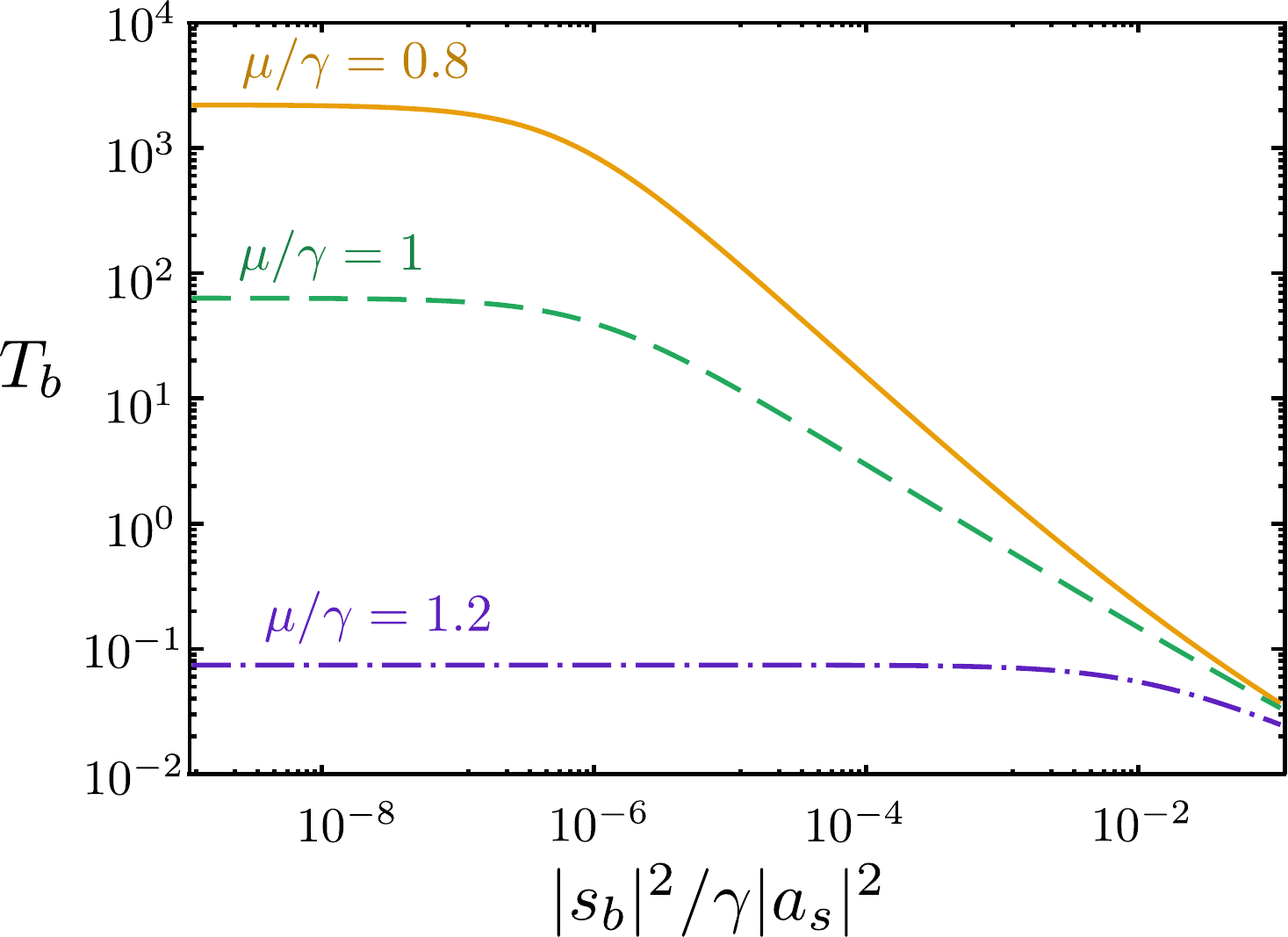}
  \caption{Transmittance $T_b$ of a back-propagating wave, versus the 
    normalized backward incident power $|s_b|^2/\gamma |a_s|^2$.  A forward-propagating wave with $|s_f|^2/\gamma |a_s|^2=10^{-4}$ is simultaneously present; the other parameters are $a_s =
    3$, $\gamma = \kappa_1 = 1$, and $\kappa_2 = 0.01\gamma$.  Note
    that $T_b$ can exceed unity because of the presence of gain in the
    system; this feature can be suppressed if desired by adding loss
    to the waveguide leads. }
  \label{fig:reflected}
\end{figure}

 Figure~\ref{fig:reflected} plots $T_b$ versus the normalized backward incident power , with fixed
forward incident power.  The features of this plot can be understood as
follows: the functioning of the isolator requires the backward wave to
cause gain saturation, but when $|s_b|^2$ is too small, the gain
saturation is dominated by the forward wave.  Thus, for small
$|s_b|^2$ we find that $T_b$ is approximately constant; in fact, it
equals the forward transmittance at the power level $|s_f|^2$.  As
$|s_b|^2$ increases, the backward wave starts to affect the gain
saturation, and isolation behavior appears in the form of a decrease
in $T_b$.  However, this is only apparent for $\mu < \gamma$
(corresponding to the $PT$-broken regime of the linear system), since
it is in this regime that the isolation ratio deviates from unity.

\section{Conclusion}

We have analyzed the relationship between the linear and nonlinear
behaviors of dual microcavity resonators with gain and loss.  The $PT$
transition of the linear system is shown to correspond closely with
the dynamical and steady-state behaviors of the gain-saturated
nonlinear system, for which $PT$ symmetry does not strictly apply.
For $\mu > \gamma$, corresponding to the linear system's
``$PT$-symmetric'' phase, the resonances are always asymptotically
stable, and the isolation ratio approaches unity.  But for $\mu <
\gamma$, corresponding to the linear system's ``$PT$-broken'' phase,
the coupled-mode dynamics become unstable at sufficiently large
frequency detunings, leading to self-sustained oscillations.  If
steady-state operation is desired, it is preferable to adopt zero
detuning.

Using the ``weak-input'' approximation, we derived a kink in the
isolation ratio at the critical point $\mu = \gamma$.  Upon relaxing
the constraint of equal waveguide-port couplings, this kink turns into
a discontinuity, meaning that the isolation ratios vary extremely
quickly with $\mu/\gamma$ in the vicinity of the critical point.  This
could be useful for using the inter-resonator coupling as an active
control parameter.  Finally, this analysis assumed that light
propagates in one direction at a time, either forward or backwards,
which holds for pulsed optical circuits where reflections appear at a
later time.  If both forward and backward waves are simultaneously
present, however, the device will fail to act as a nonlinear isolator
if the backward wave is too weak.

\appendix

\section*{Appendix A: Stability analysis}

This appendix discusses the stability analysis for the nonlinear
coupled-mode equations.  For forward transmission, we combine
Eqs.~(\ref{forward1})--(\ref{forward3}) and
(\ref{equal_loss})--(\ref{gain}) with the ``$PT$ symmetry'' condition
$g_0 = 4\gamma$, to obtain the time-dependent equations
\begin{align}
  \frac{da_1}{dt} &= \left(i\Delta\omega - \frac{\gamma + \kappa_1}{2} +
  \frac{2\gamma}{1+\left|a_1/a_s\right|^2}\right)a_1(t)
  - i\mu a_2(t), \label{time1} \\
  \frac{da_2}{dt} &= (i\Delta\omega - \frac{\gamma + \kappa_2}{2})\, a_2(t) - i\mu a_1(t)
  + \sqrt{\kappa_2}\, s_{\mathrm{in}}. \label{time2}
\end{align}
We assume a steady-state input $s_{\mathrm{in}}$, and define
\begin{align}
  a_1(t) &= \tilde{a_1} + \rho_1(t) \\
  a_2(t) &= \tilde{a_2} + \rho_2(t),
\end{align}
where $\tilde{a}_{1,2}$ is the steady-state solution that we wish to
analyze and $\rho_{1,2}(t)$ are time-dependent perturbations.  We
insert this into Eqs.~(\ref{time1})--(\ref{time2}), omitting terms
that are quadratic or higher-order in $\rho_{1}$ and $\rho_2$.  The
gain-saturation factor simplifies to:
\begin{align}
  \frac{2\gamma}{1+|a_s|^{-2} (\tilde{a}_1 + \rho_1)(\tilde{a}_1^* + \rho_1^*)}
  &\approx
  \frac{2\gamma}{1+|a_s|^{-2} (|\tilde{a}_1|^2 + \tilde{a}_1\rho_1^* +
    \tilde{a}_1^* \rho_1)} \\
  &\approx
  \frac{2\gamma}{1+\left|\tilde{a}_1/a_s\right|^2}\,
  \left[1 - \frac{\tilde{a}_1\rho_1^{*}(t)+\tilde{a_1}^{*}\rho_1(t)}
    {|a_s|^2 + |\tilde{a_1}|^2}
    \right].
\end{align}
The result is a pair of time-dependent equations,
\begin{align}
  \frac{d\rho_1}{dt} &= A\rho_1(t) + B\rho_1^{*}(t) + C\rho_2(t) \label{lyapunov1} \\
  \frac{d\rho_2}{dt} &= C\rho_1(t) + D\rho_2(t),
\end{align}
where
\begin{align}
  A &= i\Delta\omega \,-\,\frac{\gamma + \kappa_1}{2}
  \,+\, \frac{2\gamma}{1+\left|\tilde{a}_1/a_s\right|^2}
  \,-\, \frac{2\gamma\, |\tilde{a_1}/a_s|^{2}}{
    \left(1+|\tilde{a}_1/a_s|^2\right)^2} \\
  B &= -\frac{2\gamma\, \tilde{a_1}^{2} / |a_s|^2 }
  {\left(1+ |\tilde{a}_1/a_s|^2\right)^2} \\
  C &= -i\mu \\
  D &= i\Delta\omega - \frac{\gamma + \kappa_2}{2} \label{lyapunovD}
\end{align}
We then assume that the perturbations have the exponential time-dependence
\begin{align}
  \rho_1(t) &= u_1 e^{\lambda t} + v_1^{*}e^{\lambda^{*} t} \\
  \rho_2(t) &= u_2 e^{\lambda t} + v_2^{*}e^{\lambda^{*} t}.
\end{align}
Plugging these into Eqs.~(\ref{lyapunov1})--(\ref{lyapunovD}), we
derive the matrix equation
\begin{equation}
\begin{bmatrix}
    A & B & C & 0 \\
    B^{*}       & A^{*} & 0 & C^{*}  \\
    C & 0 & D & 0 \\
    0 & C^{*} & 0 & D^{*}
\end{bmatrix}
\begin{bmatrix}
  u_1  \\ v_1  \\ u_2 \\ v_2 
\end{bmatrix}
=
\lambda
\begin{bmatrix}
    u_1 \\ v_1 \\ u_2 \\ v_2 
\end{bmatrix}.
\label{lyapunov_matrix}
\end{equation}
A stable state must have Lyapunov exponents $\mathrm{Re}(\lambda) < 0$
for all four eigenvalues.  For backward transmission, we can derive
equations that have exactly the same form as
Eqs.~(\ref{lyapunov1})--(\ref{lyapunov_matrix}), except that the
steady-state amplitudes $\tilde{a}_1$ and $\tilde{a}_2$ must be
computed using Eqs.~(\ref{back1})--(\ref{back2}).

\begin{figure}
  \centering\includegraphics[width=0.95\textwidth]{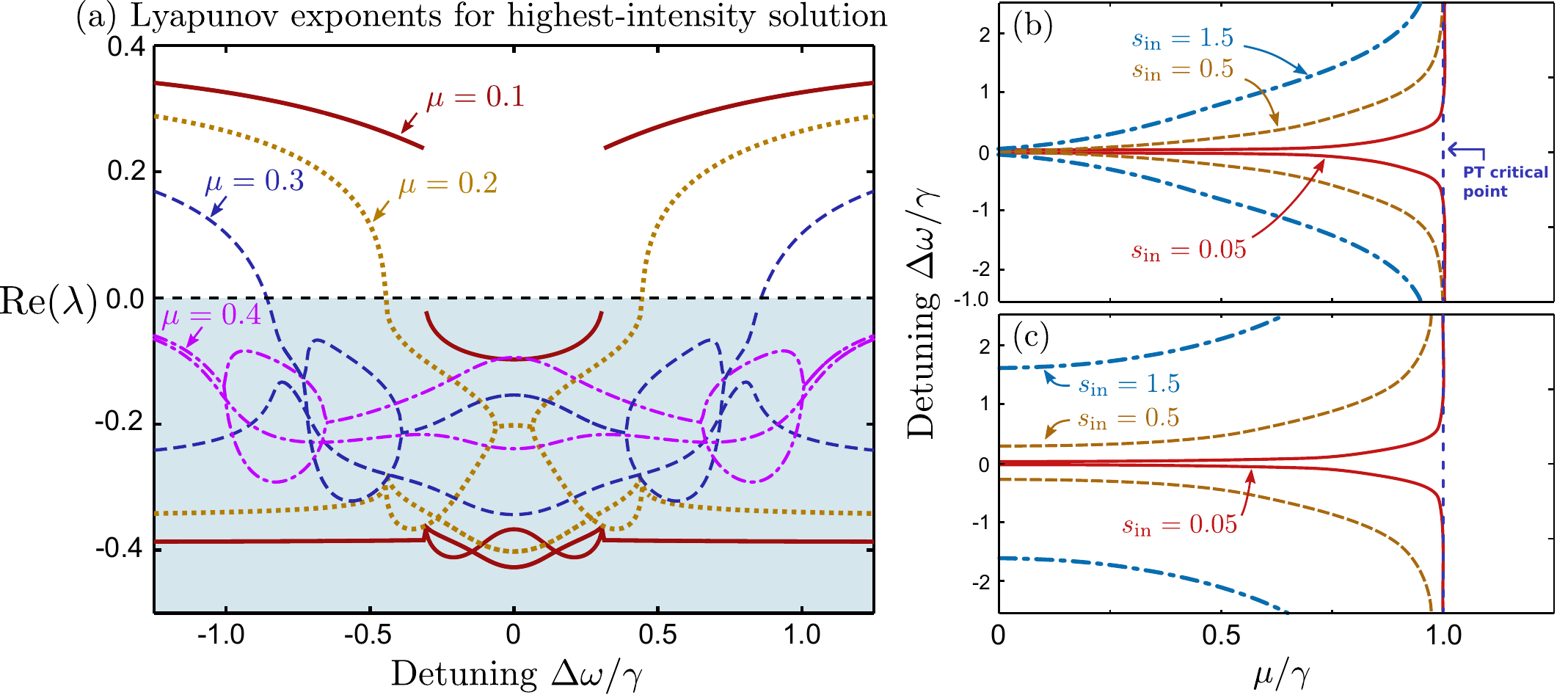}
  \caption{(a) Lyapunov exponents for the highest-intensity
    steady-state solution under forward transmission, versus detuning
    $\Delta \omega$.  Results are shown for $\mu \in \{0.1, 0.2, 0.3,
    0.4\}$.  The other model parameters are fixed at $\gamma=0.4$ and
    $s_{\mathrm{in}}=0.5$. (b) Bounds of the asymptotic stability
    region under forward transmission, for several values of the
    amplitude $s_{\mathrm{in}}$, with fixed $\gamma = 0.4$ and $a_s =
    3$.  The bandwidth of the asymptotic stability region increases
    with $\mu$, and diverges at $\mu = \gamma$, which is the $PT$
    transition point of the linear system.  (c) Bounds of the
    asymptotic stability region under backward transmission, with the
    same model parameters.  }
  \label{fig:lyapunov1}
\end{figure}

\begin{figure}
  \centering\includegraphics[width=0.95\textwidth]{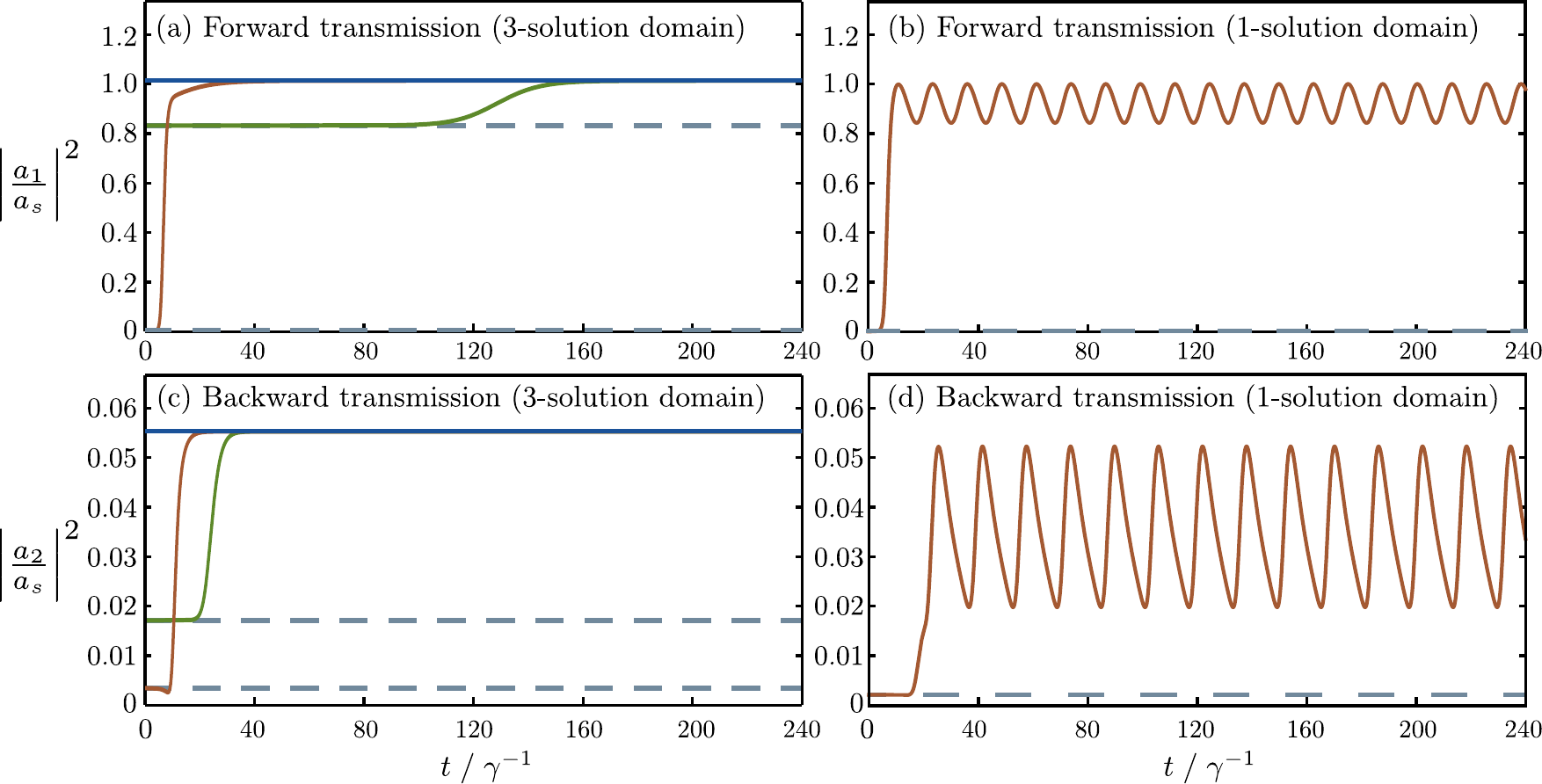}
  \caption{Time-dependent mode amplitudes under (a) forward
    transmission for $\Delta \omega = 0$ (three-solution domain), (b)
    forward transmission for $\Delta \omega = 0.2$ (one-solution
    domain), (c) backward transmission for $\Delta \omega = 0$
    (three-solution domain), and (d) backward transmission for $\Delta
    \omega = 0.2$ (one-solution domain).  The other model parameters
    are $\mu=0.1$, $\gamma = 0.4$, $a_s=3.0$, and $s_{\mathrm{in}}=0.5$.  We start
    each simulation with initial conditions perturbed from a
    steady-state solution by $\delta a_1 = \delta a_2 = 0.001$.  In
    the three-solution domain, perturbing the two lower-intensity
    solutions causes the system to evolve to the highest-intensity
    steady-state, which is asymptotically stable.  }
  \label{fig:numerical}
\end{figure}

As discussed in Section \ref{sec:multisols}, there is a domain in
parameter space where the coupled-mode equations admit three
solutions.  The Lyapunov exponents indicate that the highest-intensity
solutions are asymptotically stable.  Figure~\ref{fig:lyapunov1}(a)
plots the Lyapunov exponents for the highest-intensity solutions
versus the detuning $\Delta \omega$ (in the parts of this plot that
lie outside the three-solution domain, the highest-intensity solution
is the only one).  The exponents become negative within a frequency
band centered around $\Delta \omega = 0$.  This agrees with
Fig.~\ref{fig:multisolutions}(a)--(b).  The discontinuity in the $\mu
= 0.1$ curve results from crossing into the three-solution domain,
whereupon a new branch of asymptotically stable solutions become the
highest-intensity solutions.  For $\mu > \gamma$, the solution is
asymptotically stable for all $\Delta\omega$.  As for the
lower-intensity solutions, they are partially unstable, with one or
more exponents satisfying $\mathrm{Re}(\lambda) > 0$.

To verify these results, we solve the time-domain coupled-mode
equations numerically (using the LSODE solver).
Figure~\ref{fig:numerical}(a) and (c) shows the time-dependent
intensities, under forward and backward transmission, within the
three-solution domain ($\mu=0.1$ and $\Delta\omega=0.0$, with
$a_s=3.0$ $s_{\mathrm{in}}=0.5$ as before).  Perturbations to the
lower-intensities steady states cause the system to evolve into the
highest-intensity steady state, as expected from the stability
analysis.

\begin{figure}
  \centering\includegraphics[width=0.9\textwidth]{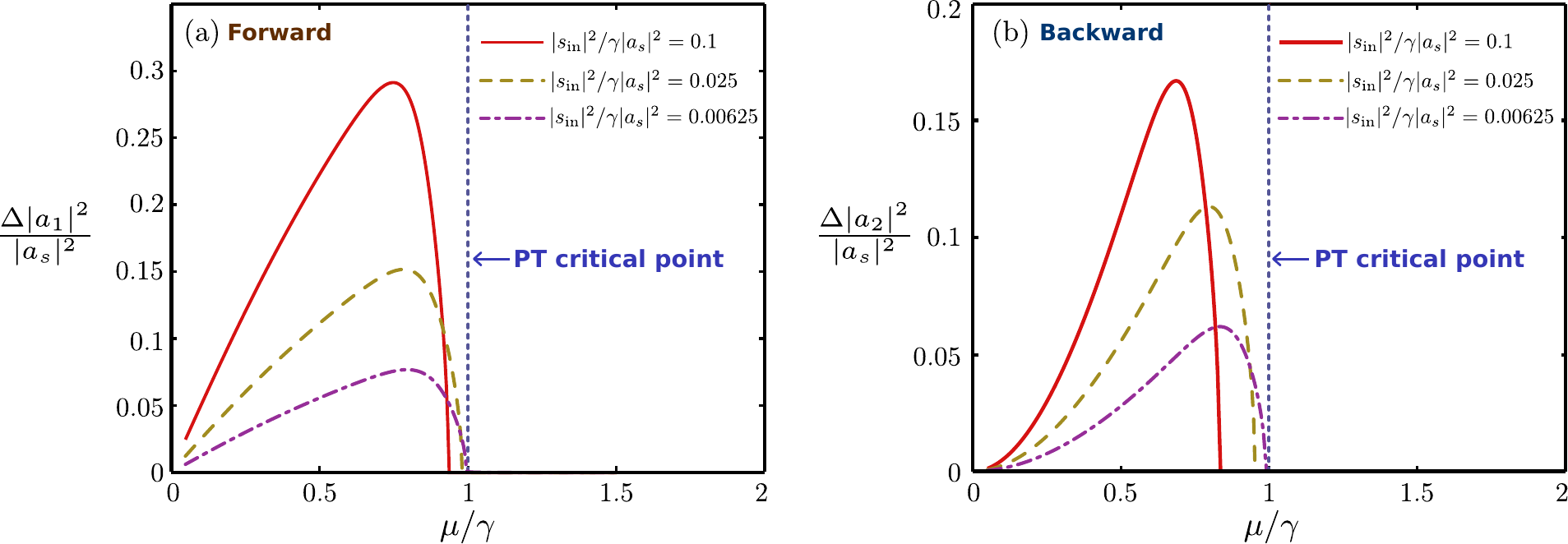}
  \caption{Beating amplitudes $\Delta|a_{1,2}|^2/|a_s|^2$, defined as the
    difference of the maximum and minimum values of $|a_{1,2}(t)|^2/|a_s|^2$
    over time $t$, versus the $PT$-breaking parameter $\mu/\gamma$.
    The amplitudes $a_{1,2}(t)$ are solved numerically using the full
    time-dependent coupled-mode equations, using $\Delta \omega =
    0.5$, $\gamma = 0.4$, $s_{\mathrm{in}}=0.5$, and $a_s = 3$. }
  \label{fig:beating}
\end{figure}

In the single-solution domain, the steady-state solution loses its
stability at large detunings, as indicated in
Fig.~\ref{fig:multisolutions}(a)--(b).  This occurs through a Hopf
bifurcation \cite{hopf_text}: small perturbations away from the steady
state induce a limit cycle, i.e.~a self-sustained oscillation in the
mode amplitudes, as shown in Fig.~\ref{fig:numerical}.  The
oscillation's mid-point coincides roughly with the real part of the
unphysical complex root of Eq.~(\ref{cubic}).

Figure~\ref{fig:beating} shows the beating amplitude versus $\mu/\gamma$.  The system is detuned so that
$\Delta\omega = 0.5$.  For small $\mu/\gamma$, the beating is
non-zero, but at $\mu \approx \gamma$ the system crosses the
asymptotic stability boundary and reaches a steady state where $\Delta
|a_{1,2}|^2/|a_s|^2 = 0$.  This is another interesting link between the
coupled-mode dynamics and the $PT$ transition.

\section*{Acknowledgments}

We are grateful to B.~Peng, H.~Wang, and D.~Leykam for helpful
discussions.  This research was supported by the Singapore National
Research Foundation under grant No.~NRFF2012-02, and by the Singapore
MOE Academic Research Fund Tier 3 grant MOE2011-T3-1-005.

\end{document}